\documentclass[%
reprint,
 amsmath,amssymb,
 aps,
pra,
]{revtex4-1}

\usepackage{graphicx}
\usepackage{dcolumn}
\usepackage{bm}
\usepackage{hyperref}
\usepackage[mathlines]{lineno}

\graphicspath{ {figures/} }
\begin{document}
\bibliographystyle{apsrev}

\title{A study of the NaI(Tl) detector response to low energy nuclear recoils and a measurement of the quenching factor in NaI(Tl)}

\author{Tyana Stiegler}
\email{tyana@physics.tamu.edu}
\affiliation{%
Texas A \& M University, Department of Physics and Astronomy,College Station, TX 77843\\
 }
\author{Clement Sofka}
\thanks{Now at: Applied Research Laboratories, The University of Texas at Austin, P.O. Box 8029, Austin, TX 78713-8029}
\affiliation{%
Texas A \& M University, Department of Physics and Astronomy,College Station, TX 77843\\
}
\author{Robert C. Webb}%
\affiliation{%
Texas A \& M University, Department of Physics and Astronomy,College Station, TX 77843\\
 }
\author{James T. White}
\thanks{deceased}%
\affiliation{%
Texas A \& M University, Department of Physics and Astronomy,College Station, TX 77843\\
 }
\date{\today}

\begin{abstract}
We analyzed the response of NaI(Tl) to low energy nuclear recoils in the experiment that is reported on here. Such detectors have been used recently to search for evidence of dark matter in the form of weakly interacting massive particles (WIMPs). Understanding the response of these detectors to low energy nuclear recoils is crucial in these searches. We have measured nuclear recoils and associated quenching factors (QF) for sodium recoils in the energy range of 8 keV - 48 keV. The results are characterized by a decrease in QF as the recoil energy decreases. The measured values are significantly lower than those reported by DAMA \cite{bernabei_2008_p4}, but are similar to results from recent measurements~\cite{collar_2013_p13,xu_2015_p34}. We present the details of our experiment, including the neutron beam calibration, shielding optimization, and the experimental design and setup. The DAMA/LIBRA combined modulation signal is used  with the new QF values to illustrate the changes to the dark matter interpretation resulting from this improved characterization of these NaI(Tl) detectors.
\end{abstract}

\maketitle


\section{\label{sec:intro}Introduction}
	Thallium doped sodium iodide (NaI(Tl)) is a popular choice of inorganic crystal scintillator. It has a high light yield and allows pulse shape discrimination between electron and nuclear recoils. The NaIAD~\cite{naiad_2005_p1}, ELEGANT-V~\cite{fushimi_1999_p2}, ANAIS~\cite{amare_2006_p3}, and DAMA/LIBRA~\cite{bernabei_2008_p4} experiments are just a few of the groups who have utilized NaI(Tl) crystals in dark matter searches. The wide range of published results from recent searches has resulted in many contradictory conclusions about the nature of possible WIMP dark matter candidates. These measurements have accentuated the importance of understanding the detector response at energies relevant to low mass WIMP searches.\par
	This experiment measured the low energy response of a single NaI(Tl) crystal using nuclear recoil response to low energy neutrons and induced radioactivity. In order to derive dark matter counting rates it is necessary to know the absolute efficiency of nuclear recoil energy to scintillation photons. This ratio is called the quenching factor (QF). To measure the quenching factor for sodium we exposed a 2-inch diameter cylindrical NaI(Tl) crystal, to a collimated mono-energetic neutron beam. Backscattered neutrons were detected between the NaI(Tl) crystal and a solid scintillating paddle detector. The paddle detector was located between the source and the NaI(Tl) crystal about one foot away from the crystal.\par 
	The following sections will discuss in detail the theoretical understanding and calculation of the QF, the scintillation response and specific characterization of our NaI(Tl) crystal, experiment, data analysis and results.
\section{\label{sec:theory}Theoretical Calculation of the Quenching Factor}
It’s well known that nuclear recoils produce fewer scintillation photons than electron recoils resulting from \(\gamma-\)ray interactions of the same energy. This effect is known as ionization quenching. After a nuclear collision, a recoiling nucleus loses energy through collisions with electrons and with other nuclei. The energy of an electron recoil that shows the same light output as the nuclear recoil is defined as the electron equivalent energy of a nuclear recoil (keV\textsubscript{ee}). The quenching factor is dependent on the rate of energy loss \(dE/dx\) and is defined as the ratio of the nuclear recoil response to the electron recoil response.
\begin{equation}
\label{eq:2p1}
QF = \frac{E\textsubscript{nr}}{E\textsubscript{er}}
\end{equation}
\par The Lindhard theory~\cite{lindhard_1963_p5,lindhard_1963_p6} attempts to quantify the energy loss from first principles in order to theoretically determine the quenching factor. The idea is to first rescale the range, or total distance, traveled by the particle inside the scintillator, \(x\), and the energy deposited, \(E_x\), of the recoiling nucleus to dimensionless variables, \(\rho\) and \(\epsilon\). The nuclear energy loss is defined as a universal function that can then be calculated numerically, with
\begin{equation}
\label{eq:2p2}
\textit{f\textsubscript{n}} (\epsilon)= (d\epsilon/d\rho)\textsubscript{n}
\end{equation}
for the nuclear energy loss and
\begin{equation}
\label{eq:2p3}
\textit{f\textsubscript{e}} (\epsilon)= \Big(\frac{d\epsilon}{d\rho}\Big)\textsubscript{e}=\kappa~ \sqrt[]{\epsilon}
\end{equation}
for electronic energy loss.\par
\begin{figure}[h]
\includegraphics[width=\columnwidth]{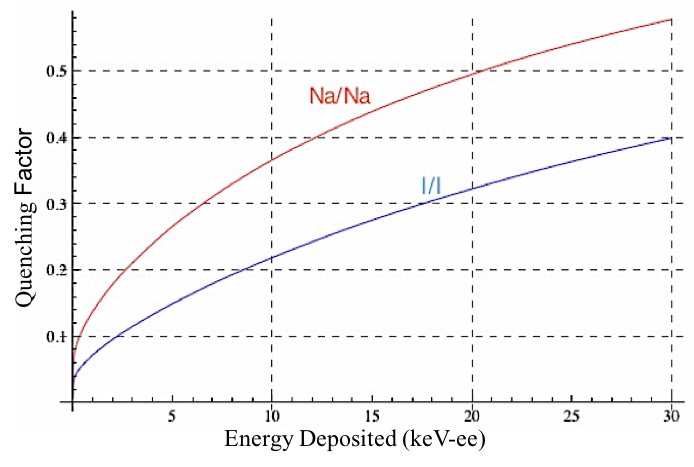}
\caption{The theoretical curves for the quenching factor of sodium and iodine recoils. The red (blue) curve represents sodium (iodine) ion recoiling onto a sodium (iodine) nucleus.}
\label{fig:lindhard}
\end{figure}

 
	When the recoiling atom is the same as the medium (for a material containing only one type of atom) then,
\begin{equation}
\label{eq:2p4}
\begin{split}
\epsilon & = \frac{11.5}{Z\textsuperscript{7/3}}E\textsubscript{x}\\
\kappa & = \frac{0.133}{\sqrt[]{A}}Z\textsuperscript{1/12}
\end{split}
\end{equation}
where \(Z\) is the atomic number of the target nuclei, \(A\) is the mass number of the target nuclei, and \(E_x\) is the energy deposited in keV~\cite{lindhard_1963_p5}. If the electronic and nuclear collisions are uncorrelated, then the total energy deposited can be split up into energy deposited into atoms, \(\eta\), and energy deposited into electrons, \(\nu\). The total energy is then 
\begin{equation}
\label{eq:2p5}
\epsilon= \eta+\nu
\end{equation} 
The electronic energy can be approximated using~\cite{lewin_1996_p7}
\begin{equation}
\label{eq:2p6}
\begin{split}
\nu & = \frac{\epsilon}{1+\kappa g(\epsilon)}\\
g(\epsilon) & = 3\epsilon\textsuperscript{0.15}+0.7\epsilon\textsuperscript{0.6}+\epsilon
\end{split}
\end{equation}
The quenching factor is then the nuclear energy deposited divided by the total energy deposited. Using Eq.~(\ref{eq:2p5}) and Eq.~(\ref{eq:2p6}) gives
\begin{equation}
\label{eq:2p7}
\frac{\eta}{\epsilon} = \frac{\kappa g(\eta)}{1+\kappa g(\eta)}
\end{equation} 
See Fig.~\ref{fig:lindhard} for a graph of Eq.~(\ref{eq:2p7}) for sodium in sodium, and iodine in iodine. The calculation of sodium in iodine is significantly more complicated and the above calculations cannot be used without modification. 
\par In semiconductors the signals from ionization agree well with the Lindhard model, which means all the energy given to electronic collisions is visible. This is not the case with solid scintillators like NaI(Tl). Measurements show that a much smaller value is observed than Eq.~(\ref{eq:2p5}) predicts. This means that there is some degree of electronic quenching as well as the nuclear quenching.~\cite{hitachi_2006_p8}\par
	Another approach to approximating the quenching factor was proposed by J.B. Birks in~\cite{birks_1964_p9}. His is a semi-empirical approach that states that the light yield of a scintillating material is dependent on the energy of the particle and the total stopping power in the material. For highly ionizing particles like protons, alphas and nuclear recoils, Birks proposes the quenching factor for ions as the light yield for ions divided by the light yield for electrons of the same energy.
\begin{equation}
\label{eq:2p8}
Q\textsubscript{i}(E)=\frac{\int_{0}^{E}\frac{dE}{1+kB(dE/dx)\textsubscript{i}}}{\int_{0}^{E} \frac{dE}{1+kB(dE/dx)\textsubscript{e}}}
\end{equation} 
where \(kB\) is defined as the Birks factor and \(dE/dx\) is the total stopping power of the ions or electrons. The Birks factor depends on experimental conditions such as temperature, amount of thallium doping, and the timing of signal collection. This makes the Birks approach difficult to compare experiment to experiment, and therefore should be used cautiously and only as a rough prediction. The Birks factor is used by Tretyak in~\cite{tretyak_2010_p10} and shows the predicted QF increasing slightly with decreasing recoil energy. An increasing quenching factor with decreasing recoil energy agrees with the recent results from~\cite{tovey_1998_p11,chagani_2008_p12}. In contrast, measurements from~\cite{collar_2013_p13} and~\cite{xu_2015_p34} show rapidly decreasing QF values in the tens of keV nuclear recoil range.
\section{\label{sec:scintResp}N\lowercase{a}I(T\lowercase{l}) Scintillation Response and Calibration}
\subsection{\label{sec:scintLight}NaI(Tl) Scintillation Light Yield}
The early studies of NaI(Tl) made it clear that the scintillation light yield was non-linear and non-proportional with respect to the energy deposited in the crystal. It was also shown that the resolution of the crystal, based on the number of scintillation photons produced, did not follow the resolution predicted by Poisson statistics. The crystal resolution is wider than prediction due to an unknown intrinsic resolution~\cite{engelkemeir_1956_p14, iredale_1961_p15, iredale_1961_p16, zerby_1961_p17}. Multiple studies have focused on quantification of the intrinsic energy resolution and measurement of the non-proportionality of the light yield for NaI(Tl)~\cite{valentine_1997_p18, rooney_1997_p19, moszynski_1997_p20, moszynski_2002_p21, khodyuk_2010_p22}. In the study by Khodyuk \textit{et al.}~\cite{khodyuk_2010_p22} the non-proportional response (nPR) and energy resolution of NaI(Tl) was measured using highly monochromatic synchrotron radiation from \(9-100\)~keV. The photon-nPR at an energy range \(E_\gamma=10-100\)~keV is measured relative to the response at \(E_\gamma=662\)~keV in percent. The shape of the photopeak-nPR is similar to other results~\cite{valentine_1997_p18}. The energy range that is of interest for this experiment is \(9-30\)~keV where the response increases from 111.5\%~-~117.2\%, a change of 5.7\%. The total energy resolution as a function of the number of photoelectrons, starts at 21.9\% and falls to 6.7\% for energies of 9~keV and 100~keV respectively.~\cite{khodyuk_2010_p22}\par
The total energy resolution of NaI(Tl) has been determined to be due to the photo-electron statistics and an additional component, termed the intrinsic resolution, which is associated with the photon-nPR~\cite{iredale_1961_p15}. There are three things associated with the light yield non-proportionality that cause the intrinsic resolution: the cascade of X-ray and Auger electrons following photoelectric absorption, the full energy absorption of \(\gamma\)-rays following multiple Compton interactions, and the statistics related to the formation of \(\gamma\)-rays~\cite{moszynski_1997_p20}.
The total resolution, \(R\), can be written as
\begin{equation}
\label{eq:3p1}
R = \sqrt[]{R^2\textsubscript{stat} + R^2\textsubscript{np}}
\end{equation} 
where \(R\textsubscript{stat}\) is the statistical resolution and \(R\textsubscript{np}\) is the intrinsic resolution. An approximate intrinsic resolution value can be calculated using a measurement of the energy resolution of a \(\gamma\)-ray of known energy, along with the statistical resolution due to the average number of photons produced at that energy. A good approximation of the statistical resolution is given by
\begin{equation}
\label{eq:3p2}
R\textsubscript{stat} = 2~\sqrt[]{2\ln2}~\sqrt[]{\frac{1+\nu}{N\textsubscript{spe}}}
\end{equation} 
where \(\nu\) is the contribution from the variance in the PMT gain (~\(\approx\)~0.25) and \(N\textsubscript{spe}\) is the average number of single photoelectrons detected~\cite{moszynski_2002_p21}. We use the electron capture peaks of \textsuperscript{128}I in Section~\ref{sec:ourCrystal} to do this calculation for our NaI(Tl) crystal.
\subsection{\label{sec:ourCrystal}Our NaI(Tl) Crystal}
In order to correctly interpret our QF measurements, we need to accurately calibrate the scintillation response of our particular NaI(Tl) crystal to electron recoils. First, we determine the number of single photoelectrons (spe) produced per keV of energy deposited. Using the number of spe's in a recoil event with a known energy we can calculate the spe/keV. For a calibration energy, we use the internal electron capture X-rays from the iodine in the crystal. The \textsuperscript{128}I becomes activated by the neutron bombardment during data taking and has a half-life of \(t_0 = 24.99\)~min~\cite{nucleardata_p23}. The combination of the low degree of activation, the short half-life, and the relatively thin lead shielding, makes it necessary to perform energy calibration immediately after the proton beam is turned off. While the beam is on, the background levels are too high to measure the electron capture X-ray signal.~\cite{nucleardata_p23}\par
\begin{figure}[!h]
\includegraphics[width=\columnwidth]{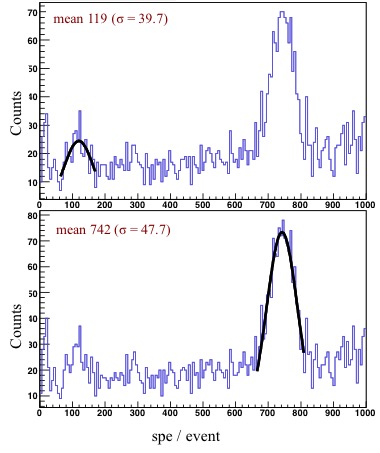}
\caption{Gaussian fit to \textsuperscript{128}I electron capture X-rays in terms of the number of single photoelectrons per event. Top: 4~keV peak, mean 118.8~spe \((\sigma = 39.7)\). Bottom: 27~keV peak, mean 741.6~spe (\(\sigma = 47.7)\).}
\label{fig:gausECap}
\end{figure}
\begin{figure}[!h]
\includegraphics[width=\columnwidth]{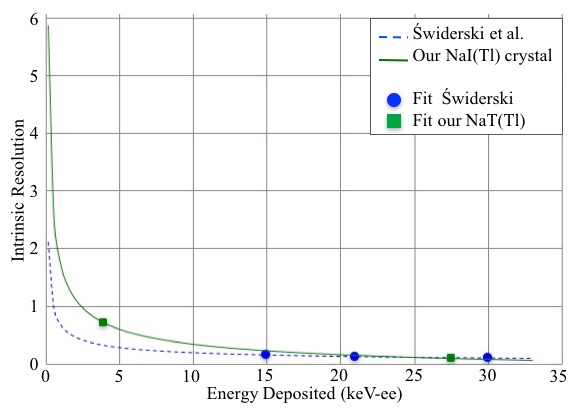}
\caption{Approximate intrinsic resolution for NaI(Tl) crystal at room temperature. Green Squares: This experiment. Blue circles: Intrinsic resolution of a NaI(Tl) crystal measured by {\'S}widerski \textit{et al.}~\cite{swiderski_2006_p24}. Green line: \(1/E^{1/2}\) approximate fit using our crystal's 4~keV and 27~keV data points. }
\label{fig:intrinsicRes}
\end{figure}
We see two Gaussian peaks that slowly decay away as time progresses. One is from the 3.77~keV and 4.03~keV lines, and the other from the 27.2~keV and 27.5~keV lines. The 4~keV peak area is approximately 8\% of the total area in the 27~keV peak.  
The time decay of the two peaks is clearly visible and a cut is made on the event number, at approximately 2 half lives, in order to maximize the peak height above background.\par
The area of the remaining events is plotted and a Gaussian fit applied to the two observed peaks (Fig.~\ref{fig:gausECap}). The standard deviation for each Gaussian fit gives a value for calculating the full width at half maximum (FWHM), 80\% for the 4~keV peak and 15\% for the 27~keV peak. The width of the observed peak is used to determine the parameters of the Gaussian spectrum that needs to be applied to the simulated gamma energies using the Geant4 framework~\cite{agostinelli_2003_p27}. The simulated X-ray spectrum uses the values from~\cite{nucleardata_p23} and applies a Gaussian spectrum to the energy of each gamma, in order to simulate the crystal response to energy deposited. The number of photoelectrons per keV of deposited energy is a Gaussian shaped curve with a FWHM that corresponds to the crystal resolution at that energy. Because each peak is made up of multiple photon energies, the measured FWHM directly from our NaI(Tl) is actually an overestimate of the true crystal resolution at each energy. The simulation uses 75\% for the X-rays near 4~keV and 14\% for those near 27~keV to produce 80\% and 15\% FWHM Gaussians at 4~keV and 27~keV respectively.

The simulation produces X-rays of the correct energies in a NaI(Tl) crystal. The resulting energy deposited values can be plotted to provide a mean kinetic energy expected for each decay peak. The mean value in spe from the detector is divided by the mean kinetic energy from the simulation to obtain the spe/keV of our crystal. We measured 30~spe/keV at 4~keV and 27~spe/keV at 27~keV. Note that the spe/keV factor for the real data is different for the 4~keV peak and 27~keV peak. This means the crystal response increases as the energy increases over this interval, as was discussed in Section~\ref{sec:scintLight}.\par
	The intrinsic resolution was calculated at 4~keV and 27~keV, then combined with other experimental results from a similar NaI(Tl) crystal~\cite{swiderski_2006_p24}, in order to estimate the approximate resolution at lower energies (See Fig.~\ref{fig:intrinsicRes}).  
\begin{figure}[t]
\includegraphics[width=\columnwidth]{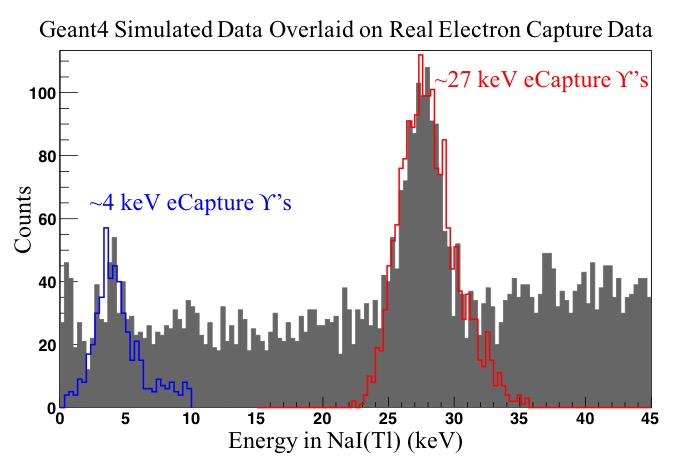}
\caption{Simulated gammas (blue and red lines) using Geant4, overlaid onto a real iodine decay data set (solid gray fill). }
\label{fig:eCapOverlay}
\end{figure}
Putting everything together, the simulated X-ray spectra are plotted with real data taken from an electron capture run Fig.~\ref{fig:eCapOverlay}. The NaI(Tl) crystal is highly activated in this particular data set, making the decay X-rays prominent. \par
\section{\label{sec:expApp}Experimental Apparatus and Procedure}
\subsection{\label{sec:setup}Detector Setup}
\begin{figure}[!hb]
\includegraphics[width=\columnwidth]{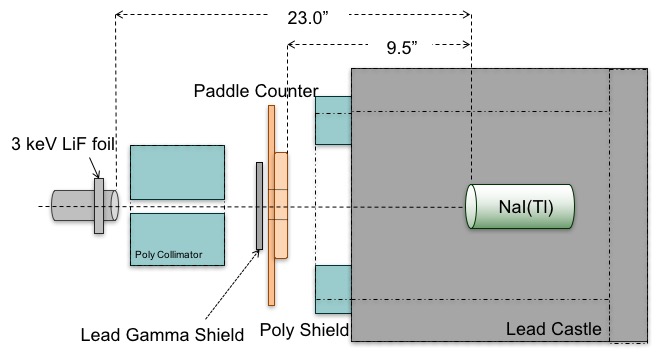}
\caption{Neutron scattering apparatus setup.}
\label{fig:detSetup}
\end{figure}
The NaI(Tl) detector setup is located at the Nuclear Science Center at Texas A\&M University, where we utilize the 2~MeV Pelletron Tandem accelerator to produce mono-energetic protons with energies up to 4~MeV. See Fig.~\ref{fig:detSetup} for a top-down view of the experiment. Neutrons are produced using the \(^{7}\)Li(p,n)\(^{7}\)Be nuclear reaction. Focused protons from the accelerator are incident on a 3~keV thick LiF coated tantalum target and the neutrons are emitted with an angular dependent energy. The resultant neutrons are incident on the collimator and shielding depicted in Fig.~\ref{fig:detSetup} to ensure a nearly mono-energetic beam reaches the 2-inch cylindrical NaI(Tl) detector. Coincidences were recorded between neutrons scattering in the NaI(Tl) and then subsequently detected in the solid scintillating paddle detector positioned to detect backscattered neutrons. In order to optimize the experimental setup a Monte Carlo simulation was performed using the Geant4 framework. The geometry in Fig.~\ref{fig:detSetup} was the final setup decided on after several different configurations were evaluated. The shielding was optimized for the highest flux of un-attenuated neutrons and lowest backgrounds from scattered neutrons and gammas.
\subsection{\label{sec:neutrons}Neutron Production and the \textsuperscript{7}Li(p,n)\textsuperscript{7}Be Reaction}
\begin{figure}[ht]
\includegraphics[width=\columnwidth]{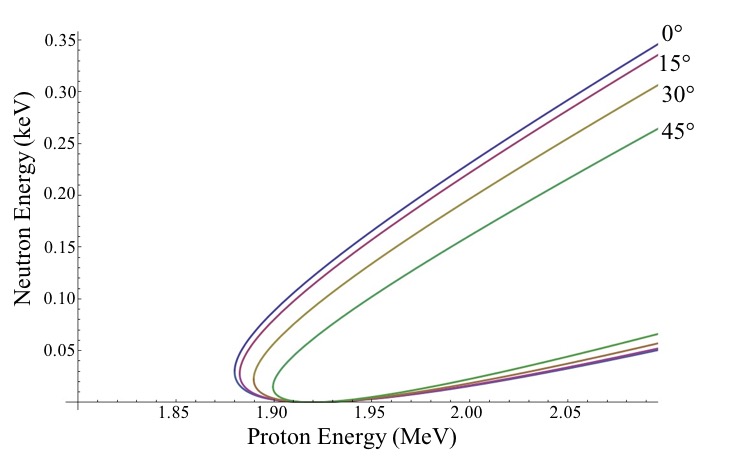}
\caption{Energy of the emitted neutrons vs incident proton energy at various emission angles near the threshold of the \textsuperscript{7}Li(p,n)\textsuperscript{7}Be reaction.}
\label{fig:protonNrg}
\end{figure}

The neutrons are produced using the \(^{7}\)Li(p,n)\(^{7}\)Be nuclear reaction and are then emitted with an energy
\begin{equation}
\label{eq:4p1}
\begin{split}
E_n = E_p\frac{m_p m_n}{(m_n+m_r)^2} \bigg\{& 2 \cos^2\theta+\zeta\delta \\
	& \pm 2\cos\theta~\sqrt[]{\cos^2\theta+\zeta\delta}\bigg\}
\end{split}
\end{equation} 
where \(\zeta\) and \(\delta\) are,
\begin{equation}
\label{eq:4p2}
\begin{split}
\zeta & = \frac{m_r(m_r+m_n)}{m_p m_n} \\
\delta & = \bigg[\frac{Q}{E_p}+\bigg(1-\frac{m_p}{m_r}\bigg)\bigg] 
\end{split}
\end{equation} 
where \(E\textsubscript{p}\) is the proton energy, \(\theta\) is the emission angle of the neutrons, \(m\textsubscript{p}\) is the mass of the proton, \(m\textsubscript{n}\) is the mass of the neutron and \(m\textsubscript{r}\) is the mass of the residual nucleus (Be). This is an endothermic reaction, with an experimental Q value of -1.644~MeV~\cite{white_1985_p25}. The reaction begins at a threshold energy given by
\begin{equation}
\label{eq:4p3}
E\textsubscript{ps} =\vert Q\vert \frac{m_p+m_t}{m_t}=1.881~MeV 
\end{equation}
where \(m\textsubscript{t}\) is the mass of the target nucleus (Li). At the threshold energy the neutrons are produced with zero energy in the center of mass frame. In the lab frame they move in a forward peaked cone with energy 
\begin{equation}
\label{eq:4p4}
E\textsubscript{ns} =E\textsubscript{ps}\frac{m_p m_n}{(m_n+m_r)^2}=30~keV 
\end{equation}
The apex angle of the emission cone is given by 
\begin{equation}
\label{eq:4p5}
\cos\theta_0 =\sqrt[]{\zeta\bigg[\frac{\vert Q\vert}{E} - \bigg(1-\frac{m_p}{m_r} \bigg)\bigg]} 
\end{equation}
Inside \(\theta\)\textsubscript{0}, at incident proton energies \(< E\textsubscript{ps}\), there are two neutron energies, corresponding to the \(\pm\) solutions in Eq.~(\ref{eq:4p1}). Each energy belongs to an emission angle in the center of mass frame. As the energy, \(E\textsubscript{p}\), of the incident proton increases, the energy of one group increases and the other decreases as the cone widens (Fig.~\ref{fig:protonNrg}). The energy of the second group goes to zero when  \(\theta_{0}\) = \(90^{\circ}\).

This is called the mono-energetic threshold energy
\begin{equation}
\label{eq:4p6}
E\textsubscript{ps} =\vert Q\vert \frac{m_r}{m_t-m_n}=1.920~MeV 
\end{equation}
For proton energies above this threshold, neutrons are produced in all directions and only the \(+\) sign holds for \(E_n\) in Eq.~(\ref{eq:4p1}).~\cite{bertulani_2007_p26}\par
The energy of the neutron beam is calibrated using the \textit{turn on} of the \(^{7}\)Li(p,n)\(^{7}\)Be reaction by observing visible counts above background in a BF\(^3\) proportional counter. The energy of the terminal potential is slowly increased until neutrons are just detected in the BF\(^3\) detector. The bending magnet reading just below this value is set as the threshold value and associated with \(E\textsubscript{ps}~=~1.881\)~MeV. The relationship between the magnitude of the magnetic field in the bending magnet and the proton energy is 
\begin{equation}
\label{eq:4p7}
E =aB^2 
\end{equation}
Once the threshold energy and magnet settings are measured, the constant can be calculated, and the energy of the subsequent events determined. Typical constant values for our experiment range from \(\textit{a} = 24-26\).
\subsection{\label{sec:nEnergy}Neutron Energy Range}
\begin{figure}[h]
\includegraphics[width=\columnwidth]{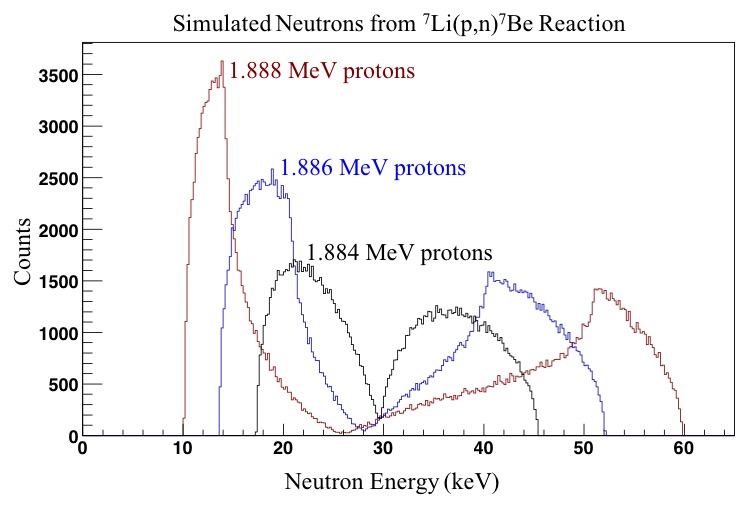}
\caption{Emitted neutron energy spectrum from the \(^{7}\)Li(p,n)\(^{7}\)Be reaction for various initial proton energies. 1.884~MeV protons (black), 1.886~MeV protons (blue), 1.888~MeV protons (red).}
\label{fig:simNeutrons}
\end{figure}

In order to accurately simulate the full spectrum of events in the NaI(Tl) crystal a Monte Carlo simulation was performed using Geant4. The neutron beam was simulated using the emission spectrum expected from Eq.~(\ref{eq:4p1}). Fig.~\ref{fig:simNeutrons} shows examples of the produced neutrons at several energy ranges near the  \(^{7}\)Li(p,n)\(^{7}\)Be threshold. The two separate energy groups that were discussed in Section~\ref{sec:neutrons} are clearly visible.\par
\begin{figure}[h]
\includegraphics[width=\columnwidth]{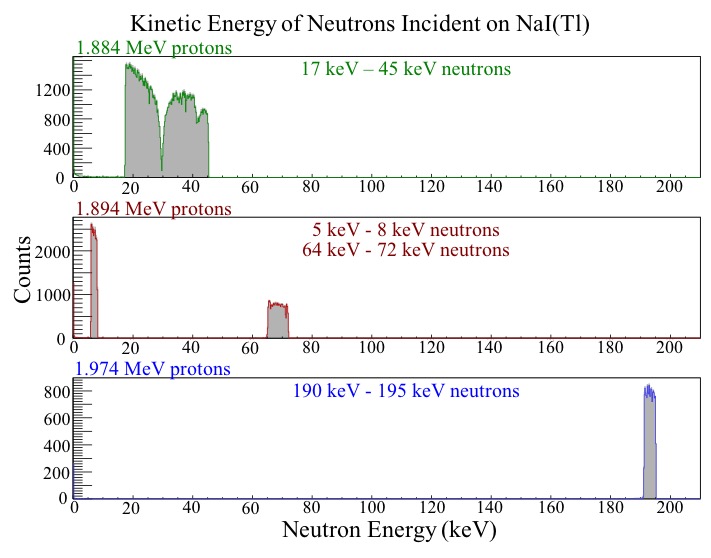}
\caption{Resultant neutrons from simulated experimental setup at three initial proton energies, above and below the mono-energetic threshold. Blue, Red and Green lines: The kinetic energies of neutrons when they reach the NaI(Tl). Dark gray fill: The energy of the neutron at the source. Top: 1.884~MeV protons. Center: 1.894~MeV protons. Bottom: 1.974~MeV protons.}
\label{fig:incidentNeutrons}
\end{figure}

The neutron energies in Fig.~\ref{fig:simNeutrons} represent the full \(180^{\circ}\) of possible emission angles. The energy of the neutron beam that actually reaches the NaI(Tl) is shown in Fig.~\ref{fig:incidentNeutrons}. The neutrons' kinetic energy at the detector, after any collisions with shielding, air molecules, or the paddle counter, are plotted over the original production energy. With this setup we attain approximately 90\% un-attenuated neutrons near threshold.\par 
Another study was performed to obtain estimates for the percent of multiple-scattered neutrons inside the NaI(Tl) crystal. This type of event is significantly reduced in our crystal due to its small size, but is still an unavoidable background. The energy profile of these events has been mapped to facilitate background cuts in the actual data. Simulations show that approximately 27\% of all events that deposit energy in the NaI(Tl) have more than one scattering of the neutron and 36\% of coincidence events between the NaI(Tl) and the paddle counter result from multi-scattering of the neutron in the crystal.\par 
\subsection{\label{sec:electronics}Electonics Hardware}
\begin{figure}[h]
\includegraphics[width=\columnwidth]{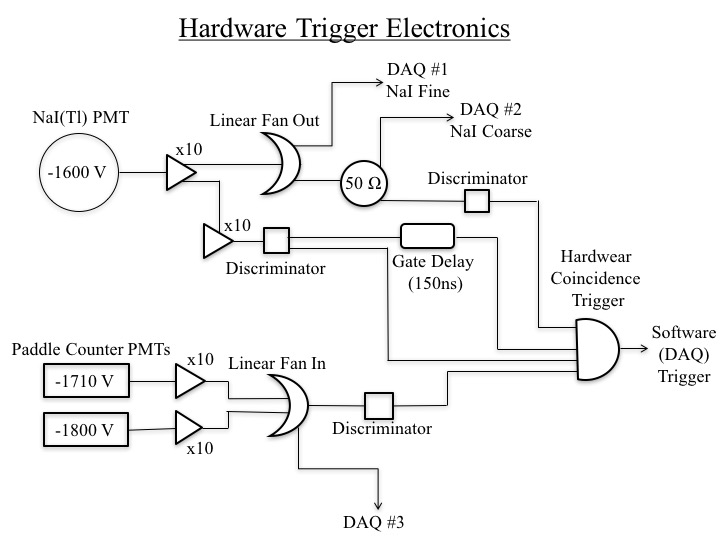}
\caption{Hardware trigger electronics chain for our experiment. Signals from the NaI(Tl) PMT and the paddle counter PMTs are amplified, split, and sent to an input channel of the DAQ and through a discriminator. The discriminator signals are sent to the external trigger of the DAQ, which is set as a coincidence between the NaI(Tl) and the paddle or as a self-coincidence in the NaI(Tl).}
\label{fig:triggerElectronics}
\end{figure}
\begin{figure}[!h]
\includegraphics[width=\columnwidth]{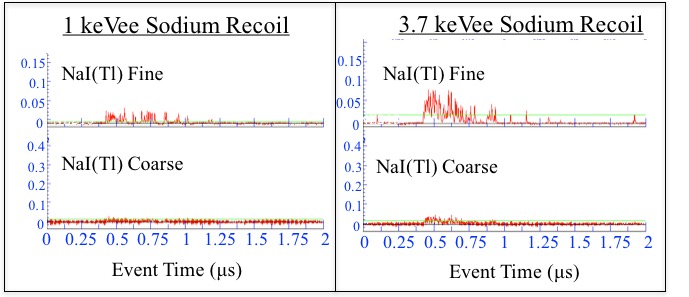}
\caption{Signals from a 1~keV\textsubscript{ee} and 3.7~keV\textsubscript{ee} nuclear recoil in the NaI(Tl) recorded with the Acqiris DC265 digitizer. Top: Fine scale NaI(Tl) channel. Bottom: Coarse scale NaI(Tl) channel.}
\label{fig:signal}
\end{figure}

Fig.~\ref{fig:triggerElectronics} shows schematically the configuration of electronics for the experiment. The signals from the paddle counter PMTs are added together and sent out to a channel of the data acquisition system (DAQ) and to a discriminator, which outputs to the coincidence trigger. The primary signal from the NaI(Tl) PMT is amplified and split, one is sent to the DAQ as the NaI(Tl) \textit{fine signal}. Another is sent through a \(50~\Omega\) splitter into a discriminator and a second DAQ channel as the NaI(Tl) \textit{coarse signal}. A third signal from the NaI(Tl) is further amplified and sent into a discriminator and a 150~ns delay generator. All three of the discriminator signals are sent to the coincidence trigger. The coincidence trigger can be used to select coincidences between the paddle and the NaI(Tl) with a 100~ns overlap, or it can be set as a self-coincidence in the NaI(Tl). The DAQ signals are recorded with an Acqiris DC265 digitizer and a 500~MHz sampling rate. Each event records a total time of \(5~\mu s\), or 2500~samples. The data acquisition software reads out the digitized waveforms and saves them for offline analysis. Fig.~\ref{fig:signal} shows digitized waveforms of nuclear recoil events in the NaI(Tl).\par
\subsection{\label{sec:events}Event Selection}
A preliminary analysis program reads each event and records the amplitude for each 2~ns sampling point. A baseline is calculated for each event using the \(50-250\)~ns window. A second set of cuts requires that no energy is deposited in the first 170~ns.\par
This second set of cuts reduces the so called \textit{tail events}, which are triggers on the low energy tails of much larger events. The tails are caused by the long decay time of NaI(Tl) scintillation pulses~\cite{birks_1964_p9} and often lead to a single event that passes the hardware trigger logic due to its coincidence with another random event in the paddle counter. Events must have 80\% of their total energy between \(170-350\)~ns and have a height to width ratio (pulse height/pulse width) \(<\)~0.1~counts/ns. These two requirements help to isolate the neutron recoils in NaI(Tl).
\section{\label{sec:data}Data Analysis and Results}
With the scintillation response of our NaI(Tl) crystal suitably characterized the quenching factor data sets can be analyzed. Starting at threshold for the \(^{7}\)Li(p,n)\(^{7}\)Be reaction and increasing in approximately \(8-10\)~keV steps to \(Ep = 2.056\)~MeV, a series of data runs were made utilizing the setup depicted in Fig.~\ref{fig:detSetup}. The NaI(Tl) and paddle coincidence trigger was used with a NaI(Tl) threshold of 0.3~keV.\par
At each energy, between 3,000 and 5,000 events were recorded and processed using the preliminary cuts discussed in Section~\ref{sec:events}. The energy deposited for each event is plotted up to 40~keV. The scale factor obtained for the 4~keV escape peak is used to scale the energy deposited into keV\textsubscript{ee} units (Table \ref{table:1}).\par
\begin{figure}[ht]
\includegraphics[width=\columnwidth]{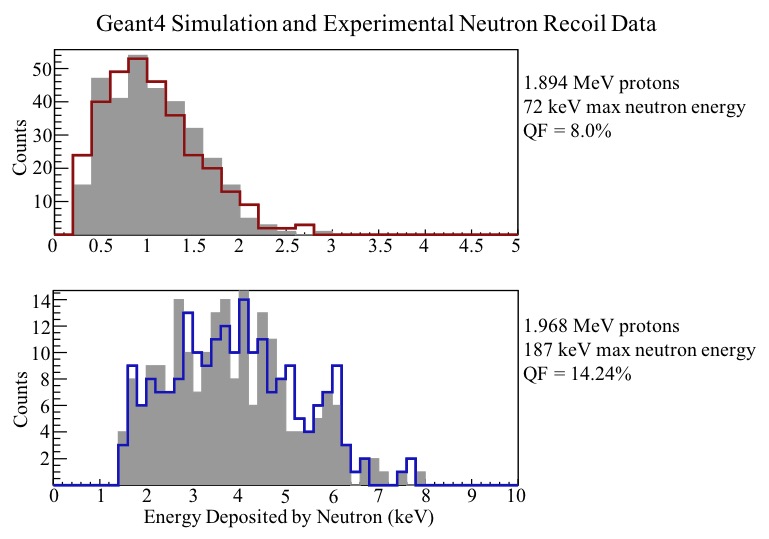}
\caption{The energy deposited in NaI(Tl) from backscattered neutrons in coincidence with the paddle counter (solid grey) with Geant4 simulated recoil spectrums overlaid (red/blue line). Top: 1.894 MeV protons. Bottom: 1.974 MeV protons.}
\label{fig:simExpOverlay}
\end{figure}
In a previous study~\cite{stiegler_2013_p29} the QF was measured using the maximum energy from \(180^{\circ}\) backscattering neutrons. That analysis benefited from a minimum reliance on the accuracy of Monte Carlo simulations for the behavior of low energy neutrons in a material. However, it suffered from additional uncertainties, due to the crystal response to intrinsic resolution, the range of neutron energies produced, the smearing of the signal from multiple scatter events, and the absence of a paddle counter for timing cuts. Each of these effects caused a smoothing out, or smearing, of the recoil spectrum away from a sharp cutoff, which resulted in a possible overestimation of the QF for the maximum neutron energy. For a more detailed discussion of these uncertainties see~\cite{collar_2013_p13,collar_2010_p30}. The goal of this experiment was to minimize or eliminate as many of these uncertainties as possible. The current analysis uses a chi-square test comparing the measured spectrum with a Monte Carlo simulation that takes into account emitted neutron energy, geometry and the total resolution of the NaI(Tl) from Section~\ref{sec:ourCrystal}.
\begin{figure}[h]
\includegraphics[width=\columnwidth]{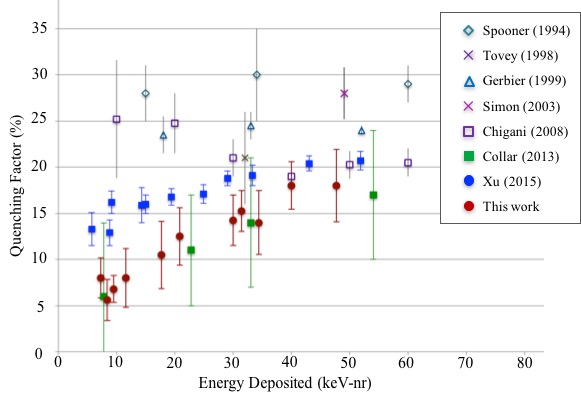}
\caption{Experimental results for the quenching factors of sodium nuclear recoils in NaI(Tl) relative to gamma rays of the same energy compared to previous published results~\cite{tovey_1998_p11,chagani_2008_p12,collar_2013_p13,spooner_1994_p31,gerbier_1999_p32,simon_2003_p33,xu_2015_p34}.}
\label{fig:results}
\end{figure}
\begin{center}
\begin{table}[h]
 \begin{tabular}{c|c|c} 
 \hline
 Observed & Simulated & Quenching\\
 Recoil Energy & Recoil Energy & Factor\\
 (keV\textsubscript{ee}) & (keV\textsubscript{nr}) & (\%)\\
 \hline\hline
 0.584 & 7.31 & 8.0 \\ 
 \hline
 0.470 & 8.39 & 5.6 \\
 \hline
 0.646 & 9.46 & 6.8 \\
 \hline
 0.928 & 11.6 & 8.0 \\
 \hline
 1.858 & 17.7 & 10.5 \\ 
 \hline 
 2.600 & 20.8 & 12.5 \\ 
 \hline
 4.275 & 30.2 & 14.25 \\ 
 \hline
 4.788 & 31.4 & 15.25 \\ 
 \hline
 4.816 & 34.4 & 14.0 \\ 
 \hline
 7.200 & 39.7 & 18.0 \\ 
 \hline
 8.586 & 47.7 & 18.0 \\ 
 \hline\hline
\end{tabular}
\caption{Experimental values for Na quenching factor. Proton energies range from 1.888 MeV - 2.014 MeV. Values plotted in Fig.~\ref{fig:results}.}
\label{table:1}
\end{table}
\end{center}
\par Geant4 is used to model the experimental apparatus shown in Fig.~\ref{fig:detSetup} and the range of incident neutron energies and trajectories from Section~\ref{sec:neutrons}. For each scattering event in the NaI(Tl) crystal, the position, energy deposited, recoiling trajectory, and interaction type is stored. The events resulting in energy being deposited in both the NaI(Tl) and paddle counter undergo a second stage of analysis where the possible quenching factor and corresponding crystal resolutions are applied to the energy deposited. The resulting simulated spectrum was used in conjunction with the experimental spectrum for the chi-square test. An example of the resulting spectrum is shown in Fig.~\ref{fig:simExpOverlay}. The quenching factor values and uncertainties are listed in Table~\ref{table:1}. The measured QF values are characterized by a decrease in value as the neutron recoil energy decreases. The single exception to this trend is the lowest energy data point corresponding to a 7.3~keV\textsubscript{nr} and 8.2\% QF. The larger measured value is most likely a result of extremely low light yield causing an artificially large QF value. See~\cite{collar_2013_p13} and~\cite{collar_2010_p30} for further discussion of this threshold effect. Our results are plotted in Fig.~\ref{fig:results} with previously published results from~\cite{tovey_1998_p11,chagani_2008_p12,collar_2013_p13,spooner_1994_p31,gerbier_1999_p32,simon_2003_p33,xu_2015_p34}.
\section{\label{sec:discussion}Discussion and Conclusions}
To more fully understand the implications of these lower QF values on dark matter limits we have used them with the DAMA/LIBRA modulation signal~\cite{bernabei_2008_p4} to determine constraints on the WIMP mass and interaction cross-sections. Standard WIMP-halo values were used and only spin-independent WIMP-nucleon interactions were considered~\cite{lewin_1996_p7,savage_2009_p35,kelso_2013_p36}. The analysis was based upon a chi-square minimization using the observed modulation from 2~keV\textsubscript{ee}\(-\)10~keV\textsubscript{ee}.\par
\begin{figure}[h]
\includegraphics[width=\columnwidth]{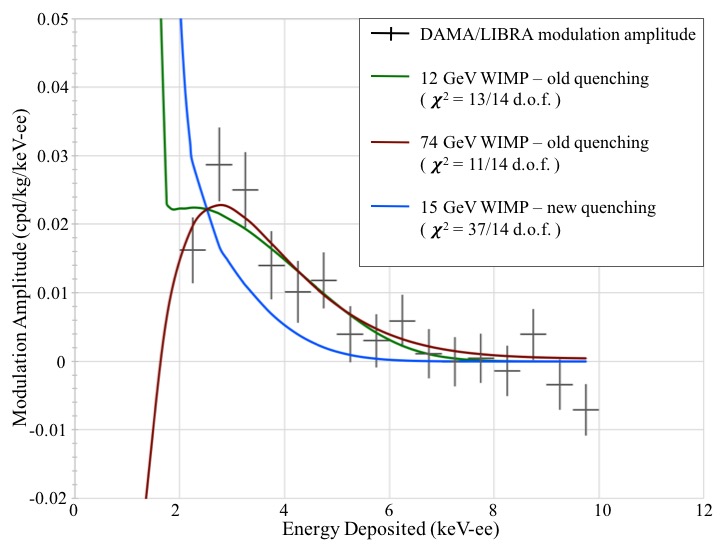}
\caption{Local best-fits for of DAMA/LIBRA modulation spectrum using the standard spin\(-\)independent WIMP model. Green and red curves use the DAMA reported Quenching factors of 0.30 for Na and 0.09 for I. The blue curve uses the new Na quenching factors for the light\(-\)WIMP fit. The heavy\(-\)WIMP fit does not change due to its dependence on the iodine quenching factor. The light\(-\)WIMP fit is much worse with the new quenching factors.}
\label{fig:modAmp}
\end{figure}
\begin{figure}[!h]
\includegraphics[width=\columnwidth]{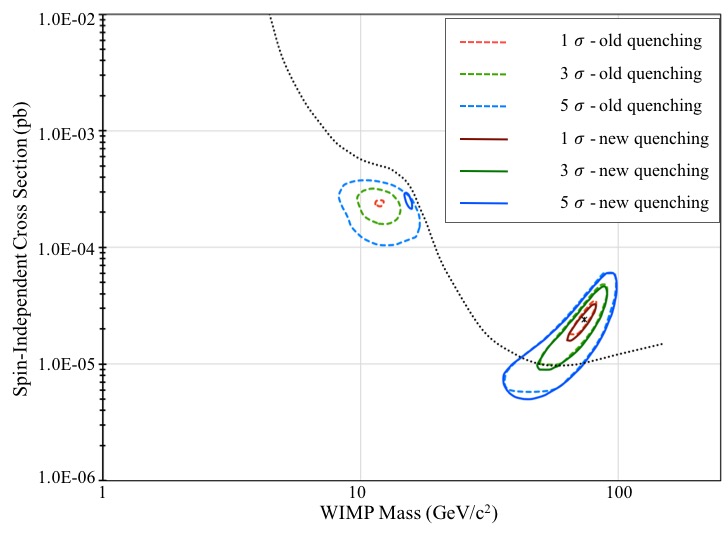}
\caption{The 1\(\sigma\), 3\(\sigma\), and 5\(\sigma\) significance contours for the DAMA/LIBRA data in the WIMP parameter space. The results using the DAMA/LIBRA quenching factors are shown in dashed lines. The results using the new quenching values are in solid lines. The heavy-WIMP contours remain unchanged while the 1\(\sigma\) and 3\(\sigma\) contours for the light-WIMP have disappeared completely. The dotted line represents the dark matter exclusion curve calculated using the overall DAMA/LIBRA observed event rate~\cite{bernabei_2008_p4}. WIMP parameters above this line are ruled out at 90\%CL.}
\label{fig:roiSIXSect}
\end{figure}
Fig.~\ref{fig:modAmp} shows the best fit WIMP signal over the DAMA/LIBRA modulation signal for both quenching factor measurements. When using the DAMA QF measurement of 0.3, a global best fit is found at a heavy WIMP mass (\(\sim\)74 GeV/c\(^2\)) with a \(\chi^2\)\textsubscript{min} of 11 with 14 degrees of freedom. There is also a local minimum at light WIMP mass (\(\sim\)12 GeV/c\(^2\)) with a \(\chi^2\)\textsubscript{min} of 13 with 14 degrees of freedom. \par 
The light-WIMP recoil is dominated by Na recoils and is therefore sensitive to the change in Na QF value. The higher mass WIMP signal is primarily due to Iodine recoils and as expected, remains unaffected in this study. The new quenching factors show that that light WIMP local minimum almost completely disappears with a new  \(\chi^2\)\textsubscript{min} of 37 with 14 degrees of freedom. The low mass WIMP region is strongly disfavored using these new quenching factor values. This results in increasing dissonance between other experiments' standard WIMP picture, and that from the DAMA/LIBRA data. The heavy-WIMP region is almost completely ruled out by the DAMA/LIBRA total rate exclusion curve, shown in Fig.~\ref{fig:roiSIXSect} as a dotted line. These regions are also ruled out by results of other experiments such as CDMSII~\cite{ahmed_2009_p37}and KIMS~\cite{kim_2012_p38}.\par
This experiment measured quenching factors for the nuclear recoils on sodium across an energy range of \(7-48\) keV\textsubscript{nr} using the coincidence between a sodium iodide detector and a scintillating paddle counter. In our earlier analysis~\cite{stiegler_2013_p29} we observed a flattening of the quenching factor value, but it lacked high counting statistics, and coincidence timing cuts. This more complete study resulted in values ranging from \(6-18\%\) QF and agrees well with the low energy measurements of~\cite{collar_2013_p13} and~\cite{xu_2015_p34}. Our current results more closely follow the Lindhard prediction curves in shape (Fig.~\ref{fig:lindhard}), rapidly decreasing in value as the recoil energy decreases. \par
The measured energy dependence of the quenching factor for sodium recoils in NaI(Tl) effects the dark matter limits previously set by other NaI(Tl) detectors~\cite{naiad_2005_p1,fushimi_1999_p2,amare_2006_p3,bernabei_2008_p4} as shown in Fig.~\ref{fig:modAmp} and Fig.~\ref{fig:roiSIXSect}. Dark matter search experiments will continue to increase the understanding of detector response and improve efficiency for filtering backgrounds in the low energy regime. The result being that limits on low mass WIMPs and their interaction cross sections will become more stringent. In order to reconcile the current, and possible future, conflicting measurements a continued emphasis should be placed on improving low-energy calibrations for dark matter detectors. 
\section*{\label{sec:acknowledgement}Aknowledgements}
This research was supported by Department of Energy Grants, DE-FG02-95ER40917 and DE-FG02-13ER42020.
\bibliography{bibTest1}

\end{document}